\documentstyle[twocolumn,epsf]{jpsj}
\def\runtitle{Randomly depleted Heisenberg ladder}
\def\runauthor{M.~{\sc Sigrist} and A.~{\sc Furusaki}}

\title{Low-Temperature Properties of the Randomly Depleted Heisenberg
Ladder}

\author{Manfred {\sc Sigrist}\footnote{On leave from Theoretische Physik,
ETH-H\"onggerberg, 8093 Z\"urich, Switzerland.} and Akira {\sc Furusaki}}

\inst{Yukawa Institute for Theoretical Physics, Kyoto University,
Kyoto 606-01}

\recdate{May 20, 1996}

\abst{
Low-temperature thermodynamic properties of the spin-$\frac12$
Heisenberg ladder system with non-magnetic impurities are discussed
using an effective Hamiltonian for the impurity-induced spins in the
background of a spin liquid with a gap.
It is shown that the uniform susceptibility shows Curie-like behavior
in high-, intermediate-, and low-temperature regimes, with three
different Curie constants. The ratio of specific heat to temperature, 
$ C/T $, is divergent in the low-temperature limit, reflecting
a singular distribution of the effective
couplings among the impurity spins. }

\kword{quasi-one-dimensional spin system, random spin chain}

\begin{document}
\sloppy
\maketitle

\begin{figure}
\begin{center} \leavevmode \epsfysize=4cm \epsfbox{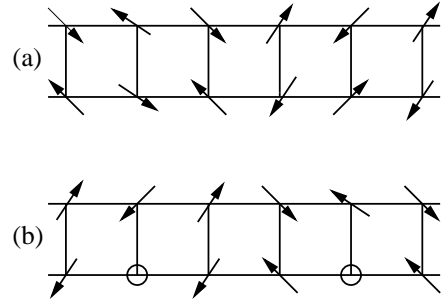} \end{center}
\caption{(a) Regular spin ladder system, (b) depleted spin ladder system.}
\label{fig1}
\end{figure}

Quasi-one-dimensional quantum spin liquids with a spin excitation gap
have attracted much interest for more than a decade. One recent
example of interest is the spin-$\frac12$ ladder system because of
its relationship to cuprate superconductors.\cite{MAURICE}
These spin ladders can be modeled by a nearest-neighbor Heisenberg
Hamiltonian of the form
\begin{equation}
{\cal H} = J \sum_i \sum_{\mu=1,2} {\mib S}_{i,\mu} \cdot
{\mib S}_{i+1,\mu} + J \sum_i {\mib S}_{i,1} \cdot {\mib S}_{i,2}
\label{H}
\end{equation}
for the case of a two-leg ladder, 
where ${\mib S}_{i,\mu}$ denotes the spin operator on rung $i$ and
leg $\mu$ (see Fig.~\ref{fig1}(a)).
Considerable effort has been invested both experimentally and
theoretically to study the properties of these systems.
We may say that by now, regular ladders, in particular, those with
two legs, are well understood both qualitatively and quantitatively.
In general ladder systems with an even number of legs have a
resonating valence bond (RVB) ground state with a spin gap, while an
odd number of legs leads to gapless excitations in a ground state with
quasi-long-range order. 
The spin gap is largest for the system with two legs, where numerical
data show $\Delta\approx 0.5J$.\cite{MAURICE}
Therefore, in real two-leg ladder systems, $\Delta$ can be very large, 
as the examples  
$ ({\rm VO})_2 {\rm P}_2 {\rm O}_7 $ with $ \Delta \sim 40 {\rm K} $ and 
$ {\rm SrCu}_2 {\rm O}_3 $ with $ \sim 400 {\rm K} $ show.\cite{MAURICE}

An interesting new aspect arises, if we take disorder into
account. The type of disorder we are interested in here, is the
depletion of spins. This may occur when ions delivering the spin
degrees of freedom are replaced at random by other ions without spins,
by accident or by intentional doping [Fig.~\ref{fig1}(b)].
For example, in ${\rm SrCu}_2{\rm O}_3$ some Cu-ions can be substituted
by nonmagnetic Zn-ions.\cite{TAKANO1,TAKANO2}
It has been observed experimentally that
in systems with a dimer ground state, such a depletion introduces a
strong component of staggered moments.\cite{EXP}
This effect has recently been investigated
theoretically.\cite{FUKU1,FUKU2,DAGOTTO} 
In the vicinity of a nonmagnetic impurity the spins develop a
strong staggered correlation, generating an effective spin doublet.  
These effective spins dominate the low-temperature
physics, since, due to the presence of the spin gap, all other
spin degrees of freedom disappear at small enough energy scales.
For a low, but finite concentration of impurities, the correlation
among these effective spins will play an important role in the
low-energy properties.
It is the aim of this letter to show some consequences of this
correlation which may be tested experimentally.

Obviously each empty site (nonmagnetic impurity) in the RVB ground
state of a two-leg spin ladder represents a spin
$S=\frac{1}{2}$.\cite{FUKU2,DAGOTTO}
In the following we will assume that the impurity concentration is
sufficiently low such that the idea of a spin-$\frac12$ degree of
freedom associated separately with each impurity is valid.
The coupling between these impurity spins is very weak and
short-ranged, because it is mediated by the spins which
participate in the formation of the RVB spin liquid state.
Before addressing the issue of coupling strengths we discuss a more
qualitative aspect.
It is intuitively clear that the coupling between the effective spins
can be either ferromagnetic or antiferromagnetic, depending
on their relative locations.
We can show this on a rigorous basis by considering a ladder system
with two impurities at sites $(i_1,\mu_1)$ and $(i_2,\mu_2)$.
The nature of the effective coupling can be determined by finding
the total spin quantum number of the ground state.
This can be straightforwardly done using Marshall's
theorem\cite{MARSHALL} which states that the spin quantum number of the
ground state of ${\cal H}$ in eq.~(\ref{H}) is identical to that of the
Hamiltonian, 
\begin{equation}
\tilde{{\cal H}} = \sum_{i \in A} \sum_{j \in B}
 {\mib S}_i \cdot {\mib S}_j ,
\end{equation}
where $ A $ and $ B $ denote the sets of sites belonging to the two
different sublattices (note that the ladder system is a bipartite
lattice). The eigenenergies $\tilde E$ of $\tilde{{\cal H}}$ depend
on three parameters: $S_{\rm tot}$ (${\mib S}_{\rm tot}=\sum_i{\mib S}_i$),
$S_A$ (${\mib S}_A = \sum_{i \in A} {\mib S}_i $) and $S_B$
(${\mib S}_B = \sum_{i \in B} {\mib S}_i $).
They are simply given by
\begin{eqnarray}
\tilde{E}(S_{\rm tot}, S_A, S_B ) &=&
 \frac{1}{2} [S_{\rm tot} (S_{\rm tot} +1) - S_A (S_A +1) \cr
&&\quad
 - S_B (S_B +1)].
\end{eqnarray}
The ground state corresponds to the case when $S_A$ and $S_B$ have
their maximal and $S_{\rm tot}$ its minimal value ($ = |S_A - S_B| $).
This immediately leads to the well-known result that if the number of
sites on the two sublattices is identical, then $S_A = S_B$ and the
ground state is a total spin singlet. 

We can now apply this theorem to our problem.
If the two impurities are on different sublattices, they remove one site
from each sublattice.
Thus the ground state again has $S_A = S_B$ and is a {\it singlet},
implying that the coupling between the two impurity spins is
antiferromagnetic.
On the other hand, if the two impurity sites lie on the same
sublattice, then the spin quantum number of this sublattice is smaller
by 1 than that of the other and $S_{\rm tot}=1$, a {\it triplet}
ground state.
This corresponds to a ferromagnetic coupling.
Consequently, the low-energy properties of the randomly depleted spin
ladder may be described effectively by a model of spins ($S=\frac12$)
with couplings of random strength and random sign. 
The low-temperature properties of this type of system have recently
been investigated by means of a real-space renormalization group
scheme\cite{WESTER,FURU} based on earlier studies by Ma and
co-workers.\cite{MA1,MA2}
In the following we discuss how these results can be applied
to this system.

Let us first consider the uniform susceptibility of this system.
At high temperatures ($k_{\rm B}T > J$) the spins are essentially free and
the susceptibility per site obeys the Curie law,
\begin{equation}
\chi(T)=\frac{\mu_{\rm B}^2}{4k_{\rm B}T},
\end{equation}
where $\mu_{\rm B}$ is the Bohr magneton and $k_{\rm B}$ is the
Boltzmann constant.
It is clear that for temperatures below the spin gap, only the
effective impurity spins contribute to the susceptibility,
because the other spins have frozen out.
This shows up as a sharp drop in the susceptibility for
$ T < \Delta $.  Since the impurity spins are only weakly coupled,
they will yield a Curie-like susceptibility in an intermediate
temperature regime:
\begin{equation}
\chi(T) = \frac{\mu^2_{\rm B} z}{4 k_{\rm B}T},
\label{chi-inter}
\end{equation}
where $z$ is the impurity density ($z\ll1$).
At lower temperatures, these impurity spins start to correlate due 
to their interaction.
First the two spins with the strongest mutual coupling freeze into either a
singlet or a triplet state, depending on the sign of the interaction.
In the case of the triplet configuration, they act as a single new spin ($
S=1 $) coupled via weaker effective interactions to its two adjacent
impurity spins.
For the singlet case, however, these spins no longer contribute
as degrees of freedom, but still mediate an effective interaction
between the two adjacent impurity spins through virtual excitations.
In this way we find the equivalent effective spin system where the
strongest bond of the original system was integrated out.
With decreasing temperature, the next strongest pair forms a single
spin with new effective coupling to its neighbors and so on.
By iteration of this process an effective spin system gradually
evolves, in which spin degrees of freedom consist of large clusters of 
randomly correlated impurity spins.
The formation of these effective spins is determined by
the energy scale $ k_{\rm B}T $; all degrees of freedom with
correlation energies larger than this energy scale are frozen into
such effective spins.
Consequently these effective spins behave essentially independently,
since the correlation energies among them are smaller than the thermal 
energy.
Using this picture we can calculate the uniform susceptibility in the 
following way.

\begin{figure}
\begin{center}\leavevmode \epsfysize=5cm \epsfbox{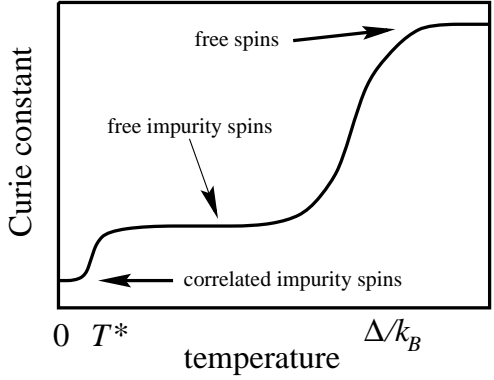}\end{center}
\caption{Schematic picture of the three regimes of the Curie constant,
$\chi T$.
The first crossover temperature is the spin gap $ \Delta $ and the
second, $ T^* $,  is related to the average coupling strength.}
\label{fig2}
\end{figure}

Let us assume that the average size of the cluster of correlated
spins contains $n$ rungs.
Note that $n$ is a monotonic function of the temperature.
Now we can estimate the average size of the effective spin associated
with the cluster by again applying Marshall's theorem.
Consider a Heisenberg ladder containing $n$ rungs where
$2zn$ spins are randomly depleted on average.
The effective spin size of the cluster corresponds to the ground-state
spin quantum number $S_{\rm tot}$ of this finite-length Heisenberg
ladder, which is, on average,
\begin{equation}
\langle S_{\rm tot} \rangle = \langle |S_A - S_B| \rangle
 = \frac12\langle | N_A - N_B | \rangle,
\end{equation}
where $N_A$ ($N_B$) is the number of depleted sites on the $A$ ($B$)
sublattice and the average is taken over the impurity configuration.
This can be easily estimated from a random-walk picture:
\begin{equation}
\langle S_{\rm tot} \rangle
= \sqrt{\frac{nz}{2}}
\label{spintot}
\end{equation}
for $ nz \gg 1 $. Thus the uniform susceptibility per site becomes
\begin{equation}
\chi(T)
= \frac{\mu^2_{\rm B}}{3 k_{\rm B}T} \frac{\langle S^2_{\rm tot}(T)
  \rangle}{2n(T)}
= \frac{\mu^2_{\rm B}z}{12 k_{\rm B}T}.
\label{chi-low}
\end{equation}
This result means that the susceptibility also follows a Curie law at
sufficiently low temperatures. However, the Curie constant for $T\to0$
is different from that in eq.~(\ref{chi-inter}).
Consequently, we expect to see crossovers in the Curie constant, as
schematically shown in Fig.~\ref{fig2}. Our prediction is that the Curie
constant drops by a factor of 1/3 from the intermediate- to the
low-temperature regime.

The characteristic temperature $ T^* $ where this crossover occurs is
set by the average coupling strength among the impurity spins 
(Fig.~\ref{fig2}).
Here we attempt to give a rough estimate of this energy scale based
on a simplified model of the spin gap state.
Suppose only two spins, at sites $(0,0)$ and $(j,\mu)$, are depleted
in the Heisenberg ladder, and the resulting impurity spins 
interact with each other by exchanging a spin triplet
excitation in the RVB spin liquid.
For calculating the long-distance behavior of this interaction, it is
sufficient to consider an approximate dispersion of these excitations.
We adopt the dispersion given by Troyer {\it et al.}, which they used to
fit numerical data of the regular two-leg spin ladder system.\cite{TROYER}
Their momentum dependence for the spin triplet excitation energy has the form
\begin{equation}
E_k = \sqrt{\Delta^2 + v^2 (k-\pi)^2},
\end{equation}
where $k$ is the momentum along the leg.
Note that this excitation corresponds to an odd-parity state along the
rung.
The effective interaction can be obtained, in a similar way to the
RKKY treatment, by second-order perturbation:
\begin{eqnarray}
\tilde{J}(j,\mu) & \approx & \frac{aJ^2}{v} (-1)^{j+\mu}
\int^{\pi}_{-\pi}{\rm d}k \frac{e^{ikj}}{\sqrt{(\Delta/v)^2+k^2}} \cr
& \approx & \frac{2aJ^2}{v}(-1)^{j+\mu} K_0 (j/ \ell_0),
\end{eqnarray}
where $a$ is an unknown constant of order unity coming from the matrix
elements.
The parameters used by Troyer {\it et al.}\ are 
$ \Delta \approx 0.5 J $ and $ v = 1.8 J $.\cite{TROYER}
Thus we obtain
\begin{equation}
\tilde{J}(j, \mu) \approx 1.1 a J (-1)^{j+\mu} K_0(j/\ell_0).
\end{equation}
It is important that the length scale appearing in the
modified Bessel function $K_0$, $\ell_0 = v/\Delta \approx 3.6$, is
rather short, and equivalent to the spin-spin correlation length in the
regular ladder.\cite{TROYER}
Therefore it seems reasonable to assume that the effective spin system 
for low-temperature physics ($T<\Delta$) is described well by
considering only nearest-neighbor Heisenberg coupling.

For a given $z$ the distribution of distance between neighboring
impurities is ${\cal P}(\ell)=2 z \exp(-2z\ell)$, which gives an
average distance $\langle\ell\rangle$ of $ 1/2z $.
This leads to an average coupling strength of
\begin{equation}
\langle|\tilde{J}|\rangle
\approx 1.1 a J \sqrt{\pi z \ell_0} e^{-1/2z\ell_0} .
\end{equation}
Thus, for $ 5 \% $ of depleted sites we get $\langle|\tilde{J}|\rangle
\sim 5 \cdot 10^{-2} a\Delta $, which could be on the order of 10K for
$\Delta\sim400$K, as in ${\rm SrCu}_2 {\rm O}_3$. 

So far we have ignored the fact that ladder systems can be
easily disconnected if we remove sites randomly. Therefore the above
analysis is valid only when a short-range
correlation among the impurity positions is present, which prohibits
configurations leading to decoupled segments of the ladder.
If we ignore this type of correlation, then the ladder
decays into decoupled segments of finite length, which,
on average, contain $1/z^2$ rungs and $2/z$ impurities.
This average length $1/z^2$ serves as a low-energy cutoff.
This means that the process of developing spin correlation with decreasing
energy scale is interrupted once all spins in a segment are frozen
into its ground-state configuration.
Hence in the low-temperature limit, each segment behaves as an
independent effective spin degree of freedom, yielding a Curie-like
temperature dependence of the susceptibility.
The Curie constant is naturally expected to be somewhere between the
values given in eqs.~(\ref{chi-inter}) and (\ref{chi-low}).
We can estimate this Curie constant by calculating the mean value of
the ground-state spin quantum number $S'$ of the segments.
For a given number $n$ of impurity spins in a segment, we obtain
\begin{eqnarray}
S'(n) & = &
\left\langle \left(\sum_{i=1}^n S^z_i \right)^2 \right\rangle^{1/2}\cr
& = & \left[ \frac{1}{2^n} \sum_{m=0}^n \frac{n!}{m!
(n-m)!} \left(\frac{n-2m}{2}\right)^2 \right]^{1/2} \cr
& = & \frac{\sqrt{n}}{2},
\label{random-walk}
\end{eqnarray}
where $ S^z_i $ denotes impurity spins $\pm 1/2$.
Noting that the distribution of $n$ is given by
$\tilde{\cal P}(n) \approx (z/2) e^{-nz/2} $, we obtain the Curie
constant
\begin{equation}
\frac{\mu^2_{\rm B} \langle S'(n) [S'(n) +1] \rangle}
     {3 k^2_{\rm B} \langle n/z\rangle}
\approx \frac{\mu^2_{\rm B}}{12 k^2_{\rm B}}
        \left(z + \sqrt{\frac{\pi}{2}} z^{3/2}\right),
\end{equation}
which gives only a small deviation from eq.~(\ref{chi-low}) for $z\ll 1$.

Next we discuss the low-temperature behavior of the specific heat.
For this purpose we need to know the distribution of the effective
couplings $\tilde J$.
The fact that the coupling strength decays exponentially
with the distance leads to a distribution which is
rather singular for $\tilde{J} \to 0$.
In this limit, it is approximated by
\begin{eqnarray}
{\cal D} (|\tilde{J}|) & \approx & \int{\rm d}\ell\, {\cal P} (\ell)
\delta(|\tilde{J}| - J_0 \sqrt{\ell_0/\ell}\, e^{-\ell / \ell_0}) \cr
& \approx & \frac{2z\ell_0}{J_0}
 \left( \frac{|\tilde{J}|}{J_0} \right)^{2z\ell_0-1},
\label{D(J)}
\end{eqnarray}
where we have neglected small logarithmic corrections. The smaller the
concentration $z$, the more of a singular distribution we obtain. 
It was shown by Westerberg and co-workers\cite{WESTER} that in a
real-space renormalization group treatment, the distribution of couplings
approaches a universal singular form\cite{note} for $\tilde{J} \to 0$.
This leads to a universal low-temperature behavior of various quantities
such as the specific heat;
at very low temperatures, the specific heat $C$ would follow a power
law, and the ratio $C(T)/T$ would be more singular than that of a
ferromagnet ($ C_{\rm ferro}/T \propto T^{-1/2}$ for $ T \to 0 $).
This can be easily seen from the following argument.

As we have already discussed, at low temperatures many impurity spins
correlate in clusters which then behave more or less as single spin
degrees of freedom.
The number of correlated impurity spins, $n(\gg1)$, at temperature $T$
can be estimated by noting that the finite-size gap, $\Delta_n$, to
the first excited state in this $n$ spin system should be on the order 
of $k_{\rm B}T$.
It is natural to assume that $\Delta_n\propto n^{-1/\alpha}$, where
the exponent $\alpha$ should be determined from ${\cal D}(|\tilde J|)$.
From eq.~(\ref{spintot}) we get
$S'(n)\propto\Delta_n^{-\alpha/2}\propto T^{-\alpha/2}$.
We can then estimate the entropy $ \sigma $ per site by assuming that each
effective spin is essentially independent:
$\sigma=z\ln[2S'(n)+1]/n$.
From these relations, we get $\sigma(T)\propto
zT^\alpha\ln(1/T)$, yielding $C(T)\propto zT^\alpha\ln(1/T)$ to
leading order in $T$.
If the couplings $\tilde J$ between the impurity spins were all
uniform and ferromagnetic, $C(T)\propto T^{1/2}$.
With random distribution of coupling, we expect that the low-lying 
excitations are, in general, softer, because they are dominated by spin 
fluctuations around the weakest couplings.
This leads to the conclusion that $ \alpha \leq 1/2 $, in agreement with
ref.~\citen{WESTER}.
Hence, $C(T)/T > T^{-1/2} $ for $ T \to 0 $. 

From these arguments we can expect the following temperature
dependence of the specific heat.
At high temperatures the specific heat behaves essentially like that of
the regular Heisenberg ladder.
It has a peak at $k_{\rm B}T\sim J$ and drops towards lower
temperature due to the presence of the spin gap.
Note that the staggered spin environment of each impurity introduces a
discrete spectrum of localized excitations in the uniform spin gap,
which can smear the gap behavior.  
As the temperature is decreased further, it shows a small 
broad peak or shoulderlike structure
around $k_{\rm B}T\approx\langle|\tilde J|\rangle$ due to the
interaction between the impurity spins, and it becomes the
anomalous power law behavior discussed above for $ T \to 0 $.
However, in the case of uncorrelated impurity positions, this
power law behavior will be cut off by the finite-size gap determined
by the average length $1/z^2$ and the coupling distribution ${\cal
D}(\tilde{J})$.
Obviously the ground state of the whole system will have a large
degeneracy, unless there is residual interaction between ``decoupled''
segments.

In summary, we have found that at low temperatures the randomly
depleted spin ladder system behaves in many respects like a random
spin system with ferro- and antiferromagnetic couplings.
Nevertheless, we emphasize that the impurities induce a staggered
correlation which is, although inhomogeneous, coherent on a long
length scale at low temperature.
This would be the dominant spin correlation feature in neutron
scattering measurements.
An important result of our study is that in this system
correlation effects lead to three distinct temperature regimes.
Of particular interest is the low-temperature regime, where Curie 
behavior of the uniform susceptibility appears with a nontrivial
Curie constant.
Furthermore, the specific heat shows an anomalous temperature
dependence for $ T \to 0 $, as long as decoupling into segments
is not significant.
These properties are not restricted to two-leg
ladders, but would also appear in any even-leg ladder and other
systems with an RVB ground state.\cite{CaVO}
We believe that for a certain range of impurity doping
in ladder systems, these effects would be in an experimentally
accessible temperature range, in particular, for two-leg
ladders with a large spin gap.
In actual Heisenberg ladder systems such as SrCu$_2$O$_3$, there is
certainly coupling between different ladders.
Although very weak, this coupling can also introduce a low-energy
cutoff for the temperature dependence of $\chi$ and $C$ and can lead
to long-range order of the staggered spin correlation, as was actually
observed experimentally.\cite{TAKANO1,TAKANO2}
This change to 3-dimensional behavior would spoil the observation of
our low-temperature regime. 

\acknowledgements
The authors thank N.\ Nagaosa, T.\ M.\ Rice and B.\ Frischmuth for
helpful discussions, and H.\ Fukuyama for providing a
preprint of his paper prior to publication. 
They are also grateful for financial support by Monbusho which made
this collaboration possible. One of the authors (M.S.)
acknowledges a fellowship (PROFIL) by the Swiss Nationalfonds.


\begin{thebibliography}{99}
\bibitem{MAURICE} See for a recent review, E.~Dagotto and T.~M.~Rice:
Science {\bf 271} (1996) 618.
\bibitem{TAKANO1} M.~Nohara, H.\ Takagi, M.\ Azuma, Y.\ Fujishiro and
M.\ Takano: preprint.
\bibitem{TAKANO2} M.\ Azuma, Y.\ Fujishiro, M.\ Takano, T.\ Ishida,
K.\ Okuda, M.\ Nohara and H.\ Takagi: preprint.
\bibitem{EXP} L.~P.~Regnault, J.~P.~Renard, G.\ Dhalenne and
A.\ Revcolevschi: Europhys.\ Lett.\ {\bf 32} (1995) 579.
\bibitem{FUKU1} H.~Fukuyama, T.~Tanimoto and M.~Saito: J.\ Phys.\ Soc.\ 
Jpn.\ {\bf 65} (1996) 1183.
\bibitem{FUKU2} H.\ Fukuyama, N.\ Nagaosa, M.\ Saito and
T.\ Tanimoto: perprint.
\bibitem{DAGOTTO} G.\ B.\ Martins, E.\ Dagotto, and J.\ A.\ Riera:
cond-mat/9605069.
\bibitem{MARSHALL} See, for example, A.~Auerbach: {\it Interacting
Electrons and Quantum Magnetism} (Springer-Verlag, New York, Berlin,
Heidelberg, 1994).
\bibitem{WESTER} E.~Westerberg, A.~Furusaki, M.~Sigrist and P.~A.~Lee:
Phys.\ Rev.\ Lett.\ {\bf 75} (1995) 4302.
\bibitem{FURU} See also A.\ Furusaki, M.\ Sigrist, E.\ Westerberg, P.\
A.\ Lee, K.\ B.\ Tanaka and N.\ Nagaosa: Phys.\ Rev.\ B {\bf 52}
(1995) 15930.
\bibitem{MA1} S.-K.\ Ma, C.\ Dasgupta and C.-K.\ Hu: Phys.\ Rev.\
Lett.\ {\bf 43} (1979) 1434.
\bibitem{MA2} C.\ Dasgupta and S.-K.\ Ma: Phys.\ Rev.\ B {\bf 22}
(1980) 1305.
\bibitem{TROYER} M.~Troyer, H.~Tsunetsugu and D.~W\"urtz: Phys.\ Rev.\ 
B {\bf 50} (1994) 13515.
\bibitem{note} For high degree of singularity in the limit $|\tilde
J|\to0$, however, we generally cannot expect that ${\cal D}(|\tilde
J|)$ approaches a universal scaling regime.
According to eq.~(\ref{D(J)}) small values of $z$ may create such
situations.
\bibitem{CaVO} A recent example may be CaV$_4$O$_9$, which shows a
plaquette RVB ground state; see K.\ Ueda, H.\ Kontani, M.\ Sigrist and 
P.\ A.\ Lee: Phys.\ Rev.\ Lett.\ {\bf 76} (1996) 1932.
\end{thebibliography}
\end{document}